# Spatial network connectivity of population and development in the USA; Implications for disease transmission

Christopher Small[1]  Andrew J. MacDonald[2]  Daniel Sousa[2]

[1] *Lamont Doherty Earth Observatory*
   *Columbia University*
   *Palisades, NY 10964  USA*

[2] *Earth Research Institute*
   *University of California Santa Barbara*
   *Santa Barbara, CA 93101  USA*

**Abstract**

The spatial distribution of population and development impose fundamental constraints on disease transmission, particularly when air travel and ground transportation are substantially reduced. Zipf's Law states that rank-size distributions of city populations follow a power law with an exponent of -1.  The assertion of a universal power law is controversial because the linearity and slope appear to vary over time and among countries.  These inconsistencies arise, in part, from administrative fragmentation of spatially contiguous agglomerations.  The issue is also complicated by the lack of any widely accepted metric for defining urban extent.  We circumvent both of these issues by treating population and development (settlements + infrastructure) as continuous fields that can be represented with measurable proxies.  In this analysis we compare census enumerations and night light luminance as proxies for population density and intensity of development in the contiguous United States. Treating population density and development intensity as continuous quantities allows for the definition of spatial networks based on the level of spatial connectivity. The resulting distributions of spatial network components (subsets of connected nodes) vary with degree of connectivity, but maintain consistent scaling over a wide range of network sizes.  The scaling properties of the component distributions provide fundamental constraints on the processes by which these networks grow and evolve.  At continental scales, spatial network rank-size distributions obtained from both population density and night light brightness are well-fit by power laws with exponents near -1 for a wide range of density and luminance thresholds.  However, the largest components ($10^4$ -$10^5$ km$^2$) are far larger than individual cities and represent spatially contiguous agglomerations of urban, suburban and periurban development.  The rank-size distributions of total population within network components are also linear over several orders of magnitude, but with somewhat greater slopes as a result of more high density urban cores occurring within larger network components. Projecting county-level numbers of confirmed cases of SARS-CoV-2 for the US onto spatial networks of population and development allows the spatiotemporal evolution of the epidemic to be quantified as propagation within networks of varying connectivity.  This suggests that using spatially explicit networks as boundary conditions for epidemiological models may yield more accurate depictions of transmission than would be obtained from using administrative divisions like counties or census tracts. The results show an abrupt transition from slow increases in confirmed cases in a small number of network components to rapid geographic dispersion to a larger number of components before mobility reductions occurred in March 2020.

**Introduction**

The spatial distribution of human population is strongly clustered, over a wide range of spatial scales. The degree of clustering changes somewhat with time and day, but generally remains within a relatively constrained extent of developed (anthropogenically modified) land area. Census enumerations indicate that the vast majority of Earth's habitable land area is very sparsely populated, and that the majority of the human population occupies only a small percentage of this area (Fig. 1). The frequently cited statistic that half of the world's population is now urban [*United_Nations*, 2018] suggests that the degree of clustering continues to increase as the largest mass migration in human history proceeds. Clearly, the spatial distribution of dense, interconnected populations is directly relevant to the transmission of many infectious diseases.

The continuum of development and habitation, as represented by night light luminance and enumerated population density, corresponds to a continuum of spatial connectivity. Densely populated, brightly lit urban cores are surrounded by less densely populated, less brightly lit urban peripheries. In some areas isolated settlements are surrounded by undeveloped land (e.g. Siberian steppe, Amazon basin, Sahara desert). In other areas settlements are embedded within a mosaic of low intensity land use often consisting of interspersed agricultural and undeveloped land. In intensive agricultural regions urban centers of different sizes can be distributed within a background of spatially contiguous agriculture. Hence, urban cores, by definition, tend to be spatially isolated while urban peripheries are more interconnected and the least developed or populated areas provide the greatest extent of spatial connectivity. The spatial structure and connectivity of population and development are clearly relevant to disease transmission, but the choropleth maps typically used to represent population and development introduce distortions unrelated to the actual distribution of people and settlements. Specifically, administrative fragmentation of spatially contiguous networks of population and development results from the use of functionally arbitrary administrative boundaries. The problem is compounded by varying definitions for different elements of the hierarchy of human settlements (e.g. [*Deuskar*, 2015]).

In this study, we circumvent the problems of administrative fragmentation and ambiguous definition by treating population density and development intensity as continuous quantities that can be depicted with measurable proxies. We use census enumerations and satellite observations of night light luminance as proxies for residential population density and intensity of development (density of lighted infrastructure) in the contiguous United States (CONUS). Treating population density and development intensity as continuous quantities allows for the use of multiple segmentation thresholds to represent varying degrees of spatial connectivity. Segmentation of a continuous field (e.g. luminance or density) creates a binary map differentiating areas with values above the threshold from areas with values below.

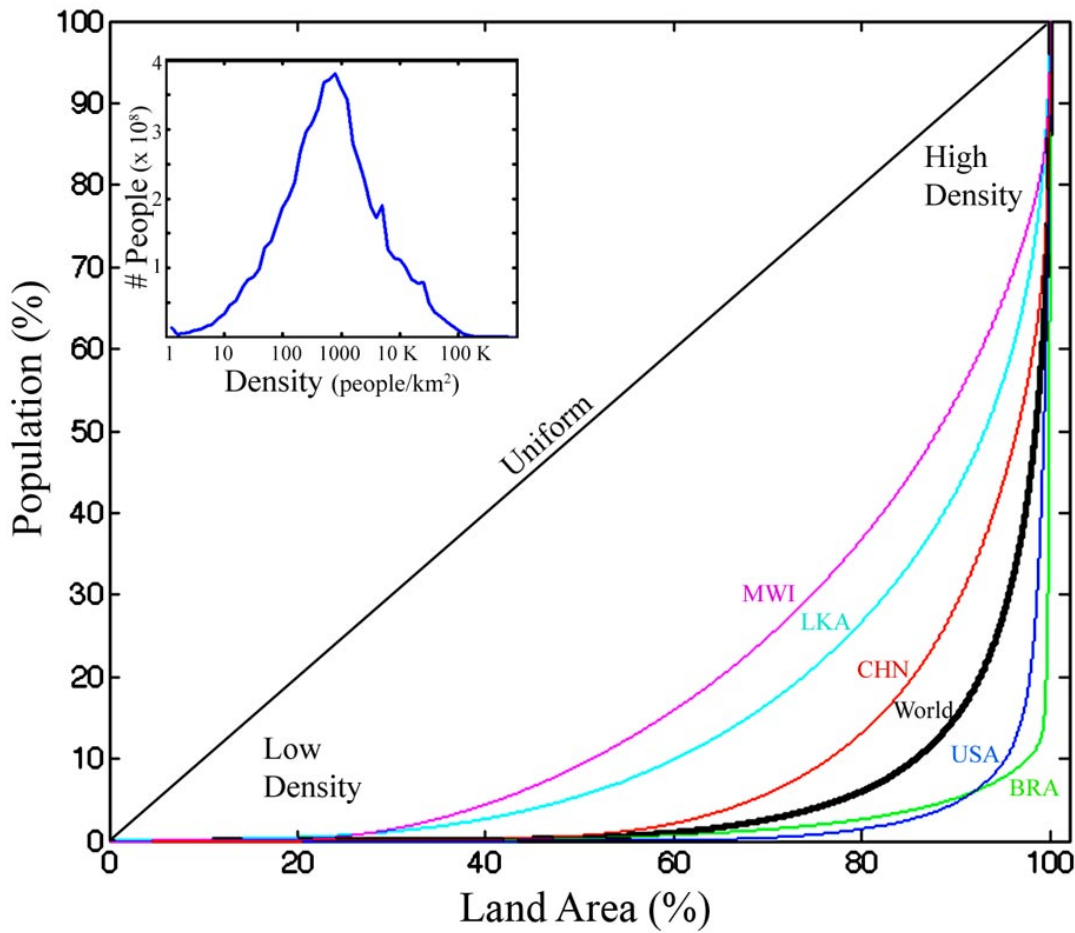

Figure 1 Global clustering of human population. Lorenz curves quantify clustering of population by degree of non-uniformity of spatial distribution on enumerated land area. Summation of census unit population counts from lowest to highest population density shows increasing degree of clustering at national level for Malawi, Sri Lanka, China, USA, and Brazil. More than half of the global population lives on less than 2% of the enumerated land area. Small settlements are not resolved in large census units so this is a conservative estimate. The global population distribution with respect to population density (inset) is somewhat more leptokurtic than lognormal.

Spatially contiguous patches (segments) of pixels or grid cells can be treated as components (connected nodes) in a spatial network. Different thresholds correspond to different degrees of connectivity within the networks. Lower thresholds represent more spatially extensive, higher connectivity networks, and vice versa. The resulting distribution of network component areas and populations provides a fundamental constraint on disease transmission through the network. We characterize the scaling properties of suites of spatial networks of population and development corresponding to different degrees of connectivity. After characterizing the population and development network distributions, we project county level numbers of confirmed cases of COVID-19 onto a more spatially detailed network of development to quantify the spatiotemporal transmission of SARS-CoV-2 detection throughout the US in early 2020. Growth of the resulting spatial network of confirmed cases within the host network is quantified in terms of number, area and host population of infected components for comparison with three metrics for mobility in March 2020 when the number of confirmed cases was increasing most rapidly.

**Background**

Host movement and interaction are crucial components of the ecology and epidemiology of pathogen transmission. Patterns of movement determine interaction networks through which pathogens are transmitted, from influenza [*Balcan et al.*, 2009] to dengue [*Reiner et al.*, 2014]. Host movement and connectivity of host populations not only influence the size and duration of local epidemics, but can also influence the extent to which transmission occurs beyond local scales, leading to larger regional epidemics [*Chen et al.*, 2018]. While such movement patterns and connectivity may be less influential to the dynamics of environmentally meditated infectious diseases—diseases with environmental reservoirs such as vectors or non-human hosts—they are critical to directly transmitted pathogens like influenza [*Balcan et al.*, 2009] and SARS-CoV-2 [*Boulos and Geraghty*, 2020]. Characterizing and modeling connectivity and movement of populations is thus critical to understanding epidemic patterns as well as to mitigating or intervening in disease spread.

Many spatial epidemiologic methods have been developed and employed to understand how patterns of movement and connectivity influence transmission, including diffusion and gravity models[*Truscott and Ferguson*, 2012], the use of commuter or air travel data to define interaction networks and connectivity [*Balcan et al.*, 2009], or metapopulation models defining interacting populations based on census-derived data [ *Balcan et al.*, 2010]. However, such census data or transportation networks are often derived from aggregate data sources, for example at the scale of administrative units like US counties or large metropolitan areas, and thus may not be representative of local scale movement and connectivity of human populations. When considering the current epidemic spread of SARS-CoV-2, particularly in the face of 'shelter-in-place' measures being taken across the globe and massive reductions in regional vehicle and air travel, such approaches are likely to be inadequate in identifying and predicting regional spread of COVID-19. Moreover, for many infectious diseases that occur in developing countries where human movement and interaction networks may be much more localized, finer scale information about population movement and connectivity patterns is imperative for improved epidemiological modeling of transmission across a diversity of geographies.

The spatial distribution of population and development (settlements + infrastructure) does show some empirical regularity in developed countries. City size distributions, defined on the basis of population, are often described as power laws. Auerbach [*Auerbach*, 1913] appears to have made the initial observation that the product of a city population and its ordinal rank is approximately constant. Lotka [*Lotka*, 1941] subsequently observed a hyperbolic rank-size relationship for U.S. city populations in 1920 and that the slope of the $Log_{10}$ rank-size plot was not exactly -1 but -0.93 and that some of the larger cities were smaller than predicted. Zipf later noted that the exponent of the power law is also close to -1 for U.S. cities [*Zipf*, 1942] as well as for the frequency of usage of words, sizes of firms and a variety of other socioeconomic characteristics [*Zipf*, 1949]. He postulated a universal principal of least effort in which the power law emerges as an optimal distribution for a variety of processes. The special case of a power law distribution with an exponent of -1 is often referred to as Zipf's Law and has been tested repeatedly for cities with mixed results [*Gabaix et al.*, 2004; *Nitsch*, 2005; *Pumain*, 2004; *Soo*, 2005]. While the consistency of Zipf's Law has driven sustained interest for several decades, and been the basis for a multitude of models [*Berry and Garrison*, 1958; *Gabaix*, 1999; *Lotka*, 1941; *Pumain*, 2006], the varying degree and extent of agreement with observation seem to preclude consensus on either the universality of the law [*Gan et al.*, 2006; *Soo*, 2005] or its underlying cause [*Batty*, 2006a; b; *Lotka*, 1941; *Pumain*, 2004; 2006]. The assertion of a universal power law for city size is controversial because the linearity and slope of the power law rank-size distribution appears to vary through time and among countries [*Gabaix et al.*, 2004; *Nitsch*, 2005; *Pumain and Moriconi-Ebrard*, 1997; *Soo*, 2005].

To some extent, the apparent inconsistency of Zipf's Law may be a result of the data and methods used to derive the parameters of the distribution. Analyses of city size distributions almost always use population counts based on census enumerations. Likewise, almost all analyses have considered city size distributions within countries or compared distributions among countries. Country-to-country differences in census collection methods and scale of administrative units make it difficult to combine data from different countries. Continent and global distributions of city size can be well fit by power laws in the upper tail but the largest cities are usually considerably smaller than predicted and the slope is generally significantly less than -1 [*Decker et al.*, 2007; *Nitsch*, 2005; *Pumain*, 2004; *Soo*, 2005]. This differs from country-specific analyses in which urban primacy sometimes results in the largest "king cities" being larger than predicted by the power law [*Jefferson*, 1939; *Pumain*, 2004]. A more pervasive complication arises in the common disparity between the functional and administrative extents of most large cities and urban agglomerations. Spatially contiguous agglomerations are generally composed of numerous smaller municipalities so a large city or agglomeration is represented not as a single entity but as number of smaller populations associated with the administrative subdivisions. For this reason, most analyses are effectively aspatial – although the function and structure of the agglomeration are clearly influenced by the aggregate spatial distribution of the population and its activities. While some studies have recognized this problem and attempted to accommodate spatial agglomerations [*Moriconi-Ebrard*, 1993; *Pumain and Moriconi-Ebrard*, 1997; *Rozenfeld et al.*, 2008], the lack of a consistent, widely accepted definition of urban extent still makes it difficult to test the universality of Zipf's Law at continental to global scales. The observation that spatial scaling of night light networks yields more consistent results than administratively defined population distributions at continent to global scales [*Small et al.*, 2011] suggests that Zipf's Law may actually be driven more by spatial network structure rather than socioeconomic factors.

**Data**

In this analysis we consider two independent, but complementary, proxies for the human habitat. 1) Spatially explicit gridded maps of population density avoid some of the complications associated with discrete administrative subdivision of urban agglomerations by depicting the presence and abundance of population as a more continuous variable (accommodating spatial discontinuities but not imposing them where not required). 2) Satellite observations of stable night light provide a unique proxy for anthropogenic development in which continuous spatial variations in luminance correspond to intensity of development as density of lighted infrastructure. Brightness and spatial extent of emitted light are correlated with population density [*Sutton et al.*, 2001], built area density [*Elvidge et al.*, 2007b] and economic activity [*Doll et al.*, 2006; *Henderson et al.*, 2009] at global scales and within specific countries.

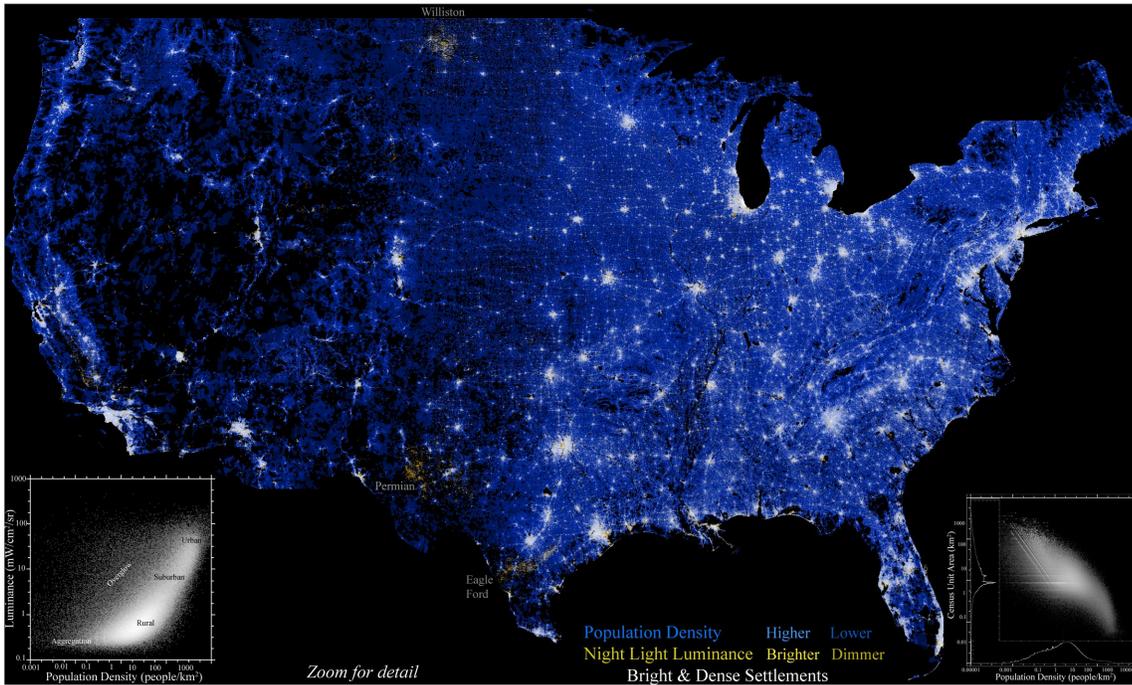

Figure 2 Night light and population of the continental United States of America. At continental scale, higher population density (blue) coincides with brighter night light (yellow) to appear white and gray. Bivariate distribution of land area (inset left) shows a correlation at higher density and luminance, but combined effects of night light overglow and census unit aggregation of sparse rural population produce a lower density tail on the distribution. Bivariate distribution of land area with density and unit area shows that higher density residential areas are enumerated in greater spatial detail. Peaks in area distribution result from large numbers of similar size counties in midwestern states. Intensive hydrocarbon production in sparsely populated rural areas appears yellow because of large numbers of lighted wells and gas flares.

The traditional population-based concept of a city can be extended to a spatially contiguous continuum of population density using geographically gridded population data. In this analysis we use the Gridded Population of the World – Version 4 (revision 11) representing global population density, circa 2010 (projected to 2015), at a spatial resolution of 30 arc seconds (~1 km at the equator) derived from ~12,500,000 enumerated administrative units with an average spatial resolution of 18 km. For the CONUS, the grid is based on ~10,500,000 census block level estimates, with an average of 12 km resolution [*Doxsey-Whitfield et al.*, 2015]. The population is distributed uniformly in space over the area of the associated administrative unit polygon using a mass-conserving interpolation algorithm [*Deichmann et al.*, 2001] so, in a sense, GPW4 represents the most uniform distribution of population possible within the constraints of the input census data. GPW4 data and documentation are available from: https://sedac.ciesin.columbia.edu/data/collection/gpw-v4.

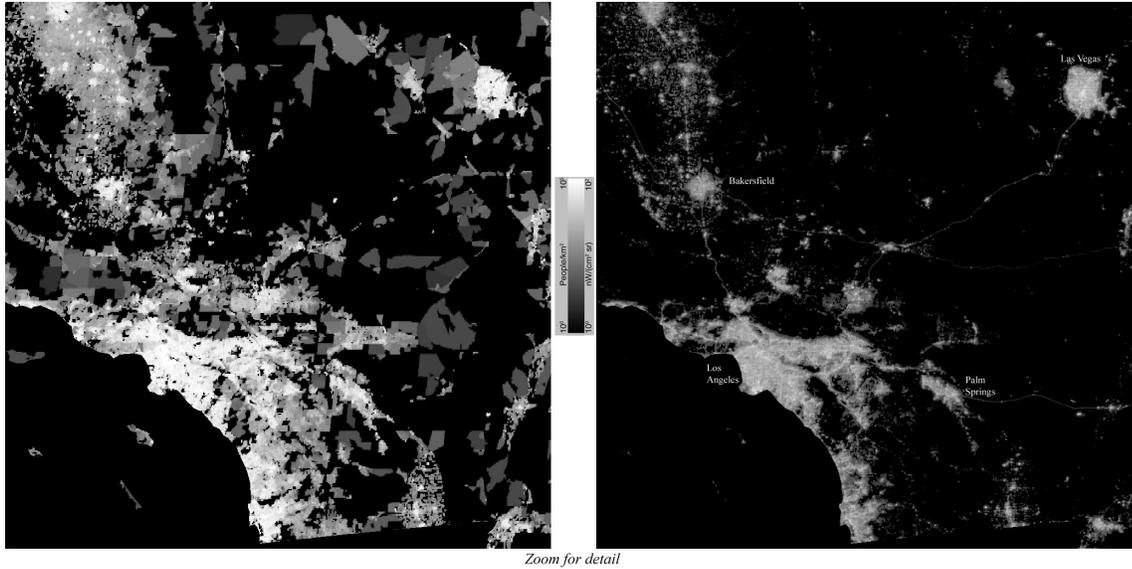
*Zoom for detail*

Figure 3  Full resolution comparison of US census population density and VIIRS night light for the southwestern USA. Census unit areas vary over 7 orders of magnitude while the VIIRS sensor maintains a constant 700 m spatial resolution. Smaller, higher density, census units in urban areas are comparable to VIIRS resolution, but the vast majority of less densely populated suburban and rural areas are mapped in far less detail.  In addition to lighted settlements, VIIRS detects light from vehicle traffic on highways, brightly lighted industrial and agricultural facilities and resource extraction.

As a complement to the population density grids we use satellite-derived observations of stable anthropogenic night light as an indication of varying degrees of development and human habitation. The Visible Infrared Imaging Radiometer Suite (VIIRS) day-night band (dnb) has been imaging night lights since 2012.  The digital data have been used to produce annual global composites of temporally stable nighttime lights from 2015 [*Elvidge et al.*, 2017]. Cloud-free annual composites of the nighttime visible band VIIRS dnb data provide average brightness in units of $nW/(cm^2\ sr)$.  Additional procedures are used to remove ephemeral lights (mostly fires) and background noise to produce gridded stable lights products.  The data and documentation are available from: https://payneinstitute.mines.edu/eog-2/viirs/.

The population density and night light brightness are coregistered at the 0.004° native resolution of the VIIRS product and reprojected to Molleweide equal area with a grid resolution of 500x500 meters.  Because both density and luminance vary over orders of magnitude, we give results as the $Log_{10}$ of each quantity, although summations of population by area are done using untransformed numbers of persons per grid cell.

**Analysis**

Rather than treat the distribution of population and development as a single canonical entity (subject to arbitrary or inconsistent definition), we acknowledge its potential sensitivity to how the quantity (e.g. settlement size or population) is defined and treat the defining criterion (e.g. brightness or density threshold) as a variable also.  This allows us to characterize the sensitivity of the network structure to the defining criterion – and quantify the effect of different degrees of spatial connectivity on the network.  By varying the lower threshold of population density (or luminance) that is part of the network, we vary the connectivity of the network.  Lower thresholds allow for larger more interconnected networks with greater numbers of components (spatially connected clusters of pixels – analogous to connected nodes). We calculate the size distribution of spatially contiguous patches (segments) above each threshold by segmenting the quantity (density or brightness) on the basis of spatial connectivity of each 0.25 km$^2$ equal area pixel to its 8 adjacent neighbors (Queen's case adjacency). We then calculate the area of each distinct contiguous segment and tabulate the size distribution for all segments (spatially contiguous patches) 9 pixels (1.5x1.5=2.25 km$^2$) or larger. Because the resulting segments represent increasingly diverse aggregations of bright or dense urban cores with dim or sparse rural areas and intermediate gradients we refer to the discrete patches as either segments, components or agglomerations, depending on the context.

Combining the density and luminance fields as a color composite (Fig. 2) highlights the widely varying resolution of the census blocks, and the coincidence of brighter light and denser population over a wide range of scales as shades of gray and white. The bivariate distribution of block size and density (inset right) shows an inverse relationship while the bivariate distribution of night light luminance and residential population density (inset left) shows a consistent scaling over 3 orders of magnitude. The latter distribution is skewed with an extended lower tail resulting from spatial dispersion of night light (overglow) and aggregation of sparse population into large census blocks. Larger, low density blocks are apparent throughout the western states as dark blue patches. Bright lights in low density blocks appear yellow. These generally correspond to industrial or resource extraction facilities. Most prominent are the densely lighted oil and gas production facilities in the Williston (ND), Eagle Ford (TX) and Permian (TX & NM) basins. A full resolution example comparison is shown in Figure 3.

For both population density and night light luminance, we treat spatial connectivity as a variable by segmenting each continuous field using different thresholds. Imposing a threshold creates a binary map distinguishing grid cells with values above and below the threshold. Each spatially contiguous patch of pixels above the threshold represents a single network component (subset of connected nodes) and the set of all components comprises the full network. We impose three thresholds on the distributions of population density and luminance (Fig. 4). For each threshold, we calculate the area and population within each component and produce rank-size distributions of component size and population. An additional low density threshold is applied to the population grid to illustrate the effect of explosive percolation that occurs as a result of the census block size distribution. With both distributions, the upper threshold admits only the upper shoulder containing the brightest (densest) pixels corresponding to urban cores. Progressively lower thresholds are added until the background luminance (or aggregate density) is reached, at which point the ability of the data to represent the extent of development (residential population) becomes questionable. In addition to providing a proxy for spatial connectivity, comparing multiple thresholds allows us to test the sensitivity of the resulting network structure to the threshold chosen.

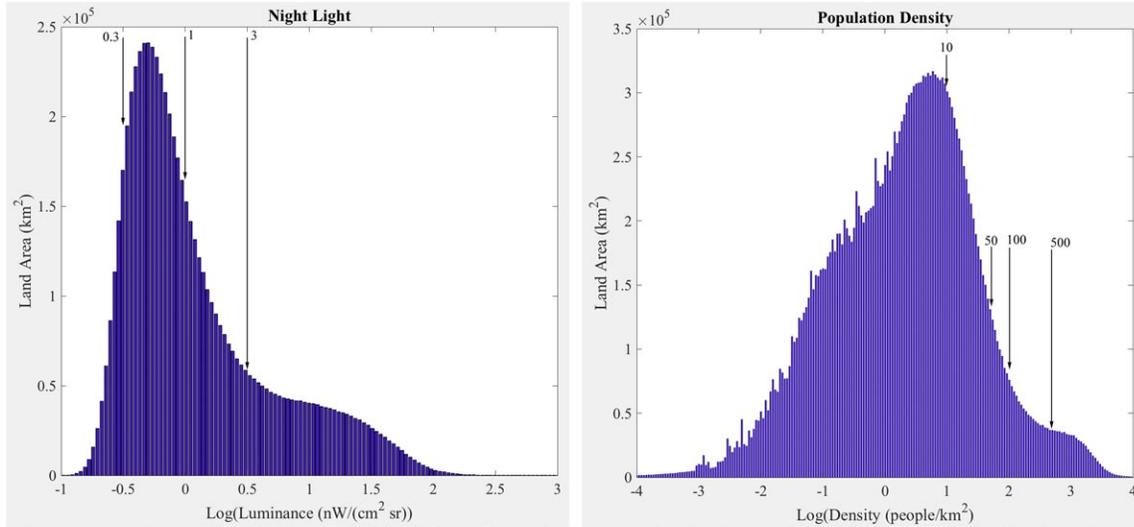

Figure 4 Land area distributions of night light and population density for the continental US. Both distributions have a prominent shoulder in the upper tail corresponding to more densely populated, brightly lighted urban and suburban areas. The night light distribution also has a much larger mode associated with vast areas of dark land surface (not shown). Threshold levels shown for both distributions are the lower cutoffs used to segment the spatial luminance and density fields into networks of spatially contiguous components (connected pixel clusters) brighter/denser than the threshold.

We calculate two parameters of the rank-size distributions of area and population (within area above threshold): domain and slope (or exponent). Previous analyses of city population size distributions [*Rosen and Resnick*, 1980; *Soo*, 2005] and other purported power laws [*Clauset et al.*, 2009; *Newman*, 2005] indicate that the estimate of the exponent can be sensitive to biases inherent in the method of estimation. In this study we use the Maximum Likelihood Estimate (MLE) for a power law to quantify the distribution of segment sizes. Although the Ordinary Least Squares (OLS) approach is, by far, the more commonly used method for estimation of Zipf exponents, it suffers from a number of shortcomings as a means of quantifying and testing the power law hypothesis [*Clauset et al.*, 2009; *Newman*, 2005; *Sornette*, 2003]. We use more statistically sound estimates of the power law exponent and optimal rank-size cutoff, derived using the MLE and semi-parametric bootstrap approach given by Clauset and colleagues [*Clauset et al.*, 2009].

**Results**

The spatial networks of population and development that result from each threshold are compared for the CONUS in Figure 5a, with full resolution comparisons in Figure 5b. As expected, lower thresholds produce greater numbers of network components spanning a wider range of sizes. The increase in component size illustrates the effect of greater network connectivity with lower threshold values. The full resolution comparisons in Figure 5b illustrate the greater detail resolved by the night light which extends the lower threshold network into periurban peripheries where larger census blocks do not resolve smaller individual settlements. The inset map on Figure 5a shows the effect of an additional lower threshold on the population density network. As the threshold approaches the mode of the density distribution, greater numbers of larger, lower density census blocks exceed the threshold and interconnect many of the larger components in the eastern US to form a single

Giant Component encompassing a much larger area than any of the others. This is a result of the underlying census block size distribution, rather than a characteristic of the actual population density. The result is analogous to the process of explosive percolation, which is commonly observed with critical phenomena (e.g. [*D'Souza et al.*, 2019]).

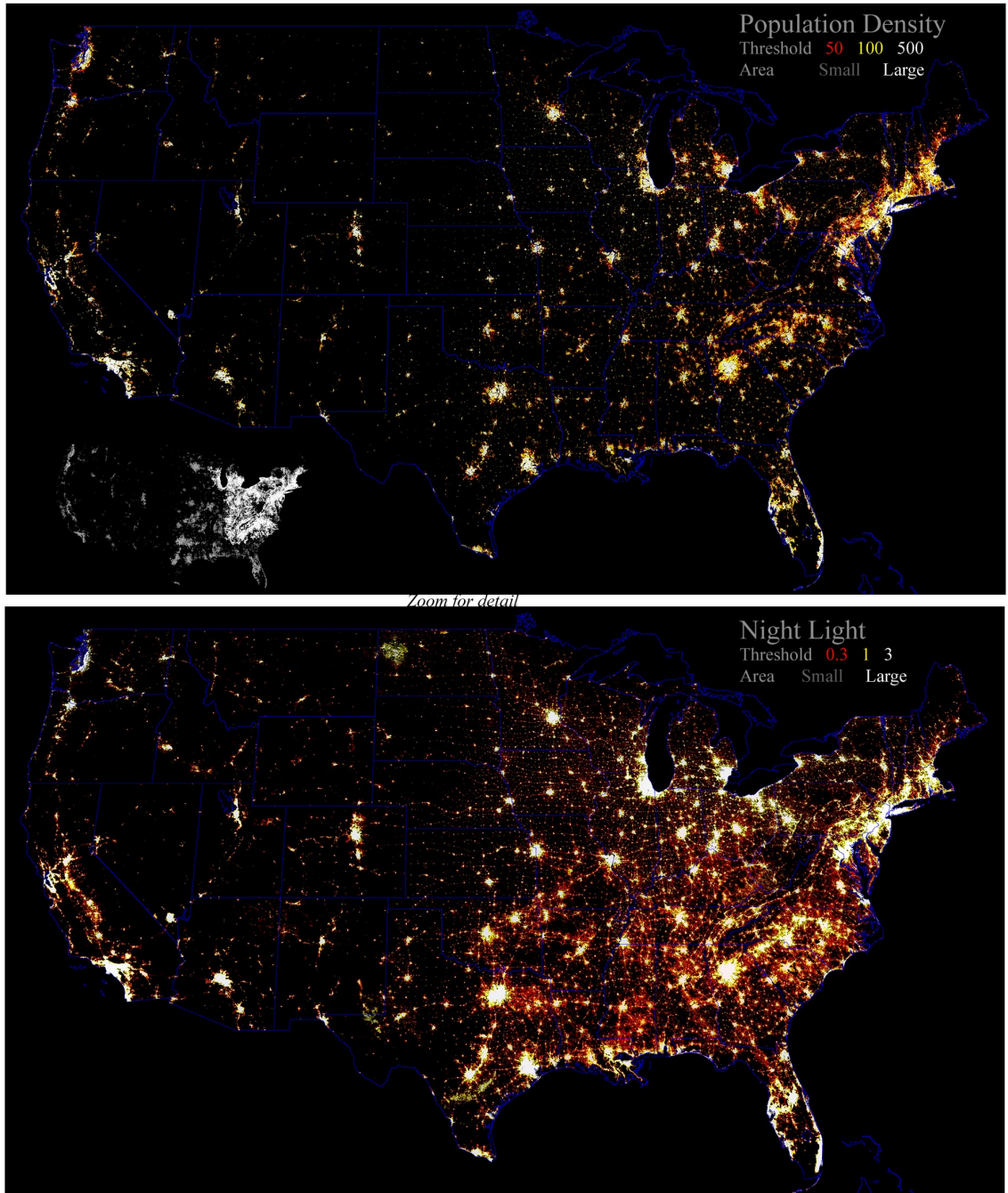

Figure 5a Spatial network extents of population density and night light. For both density and luminance fields, lower thresholds produce greater numbers of components of all sizes with increasing degrees of connectivity. However, the night light networks are larger and more interconnected than the population networks at these densities. The inset map shows the network produced by the lowest density threshold (10 people/km$^2$). At this threshold, explosive percolation produces a giant component that encompasses most of the eastern US.

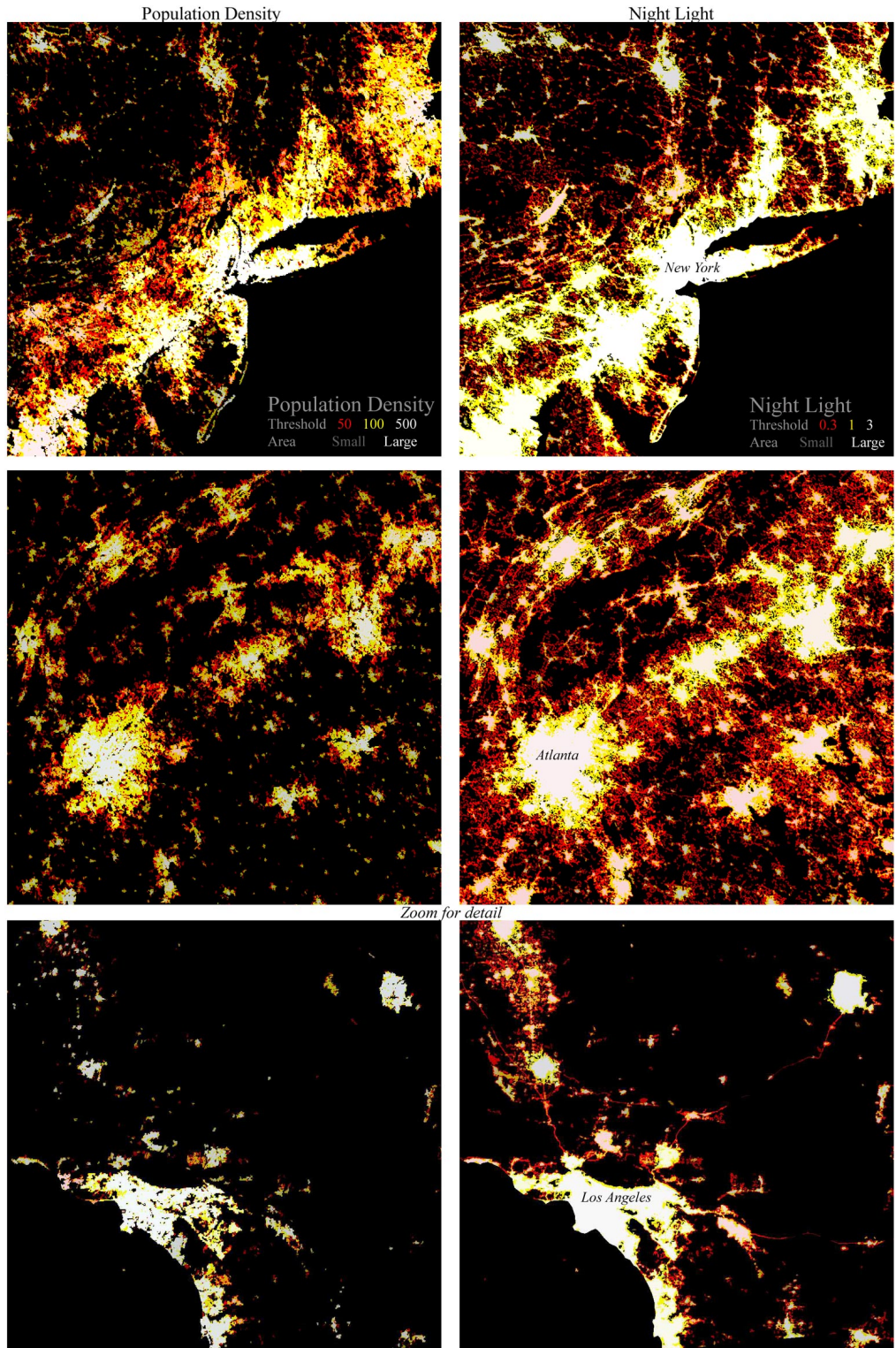

Figure 5b Spatial network extents of population density and night light. For both density and luminance fields, lower thresholds produce greater numbers of components of all sizes with increasing degrees of connectivity. However, the night light networks are larger and more interconnected than the population networks at these densities. Density networks are more detailed in urban cores, but light networks are more detailed elsewhere.

The component size distributions of spatial networks of lighted development show strongly linear rank-size distributions, consistent with a power law scaling (Fig. 6). Increasing from lower to higher thresholds has the expected effect of reducing both the size and number of components, but the slopes remain very near -1. The rank-size distributions of total component populations are also strongly linear over 4 and 5 orders of magnitude in number and size (respectively), although the slope is greater than -1. The roll off of population in the lower tails of the distributions reveal the point where the largest block units become much larger than the 500m grid resolution and population density drops precipitously. In each case, the optimal MLE fit corresponds to a cutoff in the lower tail of the distribution, near the point of downward curvature in the case of the population distributions.

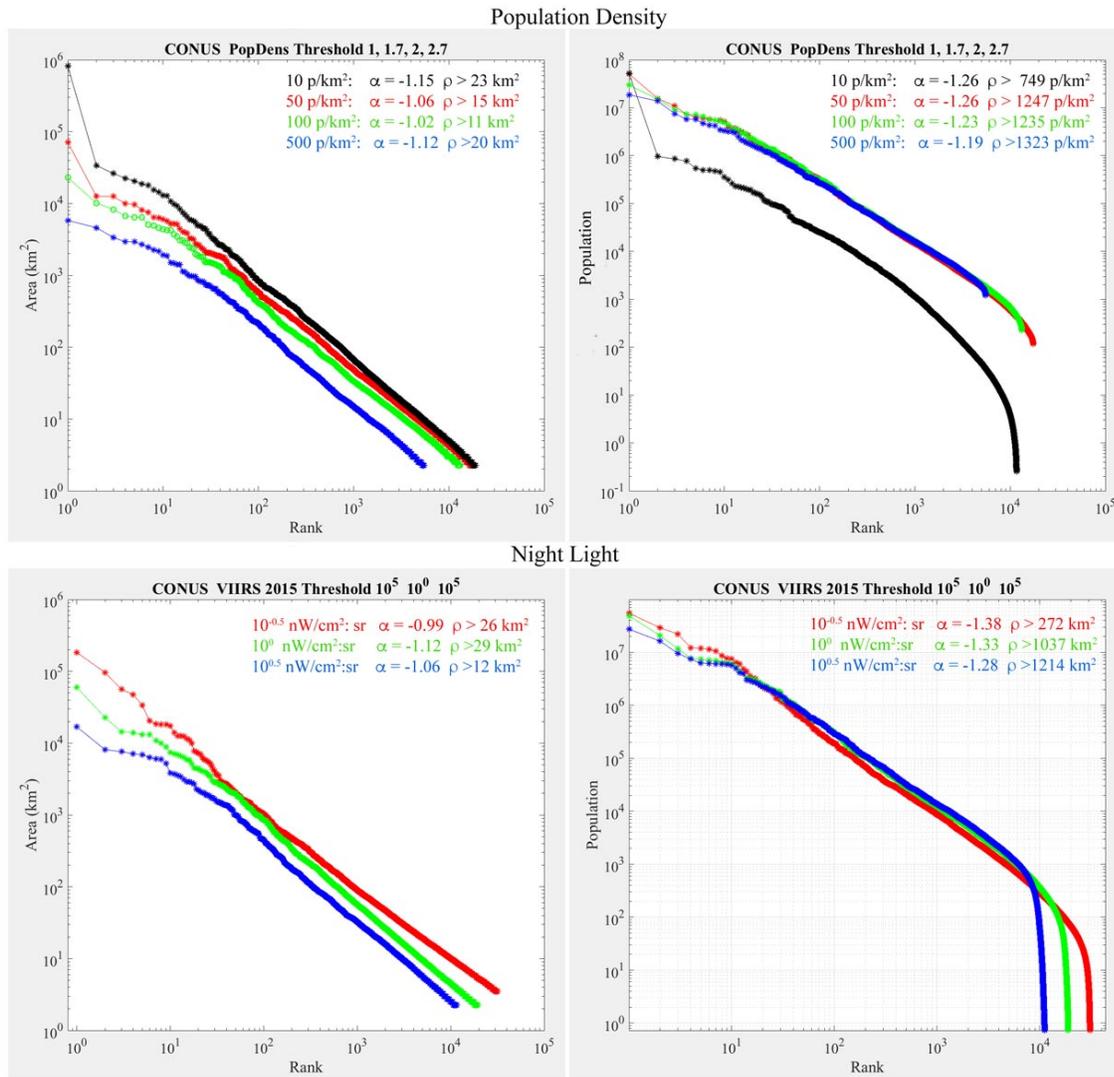

Figure 6 Rank-size distributions for population density and night light networks. Both population density and night light show strongly linear scaling spanning 4 to 5 orders of magnitude in component area and population. Network area distributions show expected variation in number and area of components, with slopes near -1 for all thresholds. Network population distributions are threshold-invariant, reflecting a complementary scaling of density distribution with network area. The progressive growth of the largest component in the population area network results in explosive percolation at the lowest threshold, with the emergence of a giant component containing ~75% of the total population.

The spatial networks obtained from population density show very similar scaling to those derived from night light (Fig. 6). Both linearity, range and slopes are very similar, indicating the degree to which a wide range of densities and brightness are spatially correlated. One important difference arises in the growth of the largest (Giant) component at lower density thresholds. This occurs as the result of extensive areas of lower mean population density in the eastern CONUS resulting in superconnection of the many already large components into a single very large Giant Component. This is analogous to explosive percolation observed in critical phenomena.

It is noteworthy that network component area scaling maintains slopes near -1 for both population density and luminance, but the distributions of total population within components do not. In part, this is a result of the greater frequency of high density urban cores within larger network components, and the tendency of smaller settlements to have lower densities. Given the near-unity scaling of the network area distributions, a skewed distribution of population density within these networks is expected to be manifest in greater slopes of corresponding population size distributions.

The implications of this spatial network scaling for disease transmission are illustrated by projecting county level numbers of confirmed cases of SARS-CoV-2 onto the networks derived from night light luminance. Spatially intersecting the low threshold/high connectivity ($10^{-0.5}$) segmentation of $Log_{10}$(Luminance) with the spatial extent of counties reporting confirmed cases produces a subnetwork of night light components within and containing reporting counties. The intersection reflects the spatial distribution of development within each county, thereby showing specific areas where people are more likely to be. Adding the $Log_{10}$(cases) to $Log_{10}$(population) at grid resolution highlights higher density areas with larger numbers of cases. This could be considered a proxy for potentially exposed population projected onto a more detailed map of likely population distribution (developed areas). Figure 7 shows this combined product (red) against a background of counties not reporting confirmed cases (green) on three dates in March 2020. It is clear that the counties reporting cases spread throughout almost the entire network within the month of March. Time-lapse cartography of the spatiotemporal evolution of the daily confirmed case counts clearly depicts the slow growth of cases in a small number of network components on the west coast, followed by the appearance of isolated clusters in Chicago and Boston, with an abrupt jump to several other network components in the eastern US in the second week of March (*https://youtu.be/DA-U5XCqLO4*).

Rank-size distributions of network components with confirmed cases clearly depict the propagation through the network. Figure 8 shows the evolution of the distributions for both high and low connectivity networks over the month of March. A small but increasing number of large components grows much faster than the larger number of smaller components, and the sigmoid nonlinearity propagates from larger to smaller components, continuing to increase the area of the network as it propagates to larger numbers of smaller, more remote components. Distributions of network areas and populations behave similarly, with almost the entire population potentially exposed, despite large areas with no confirmed cases. This reflects both the large fraction of the population living in small, high density areas and the increasing underestimation of the true population in the lower tail of the distribution.

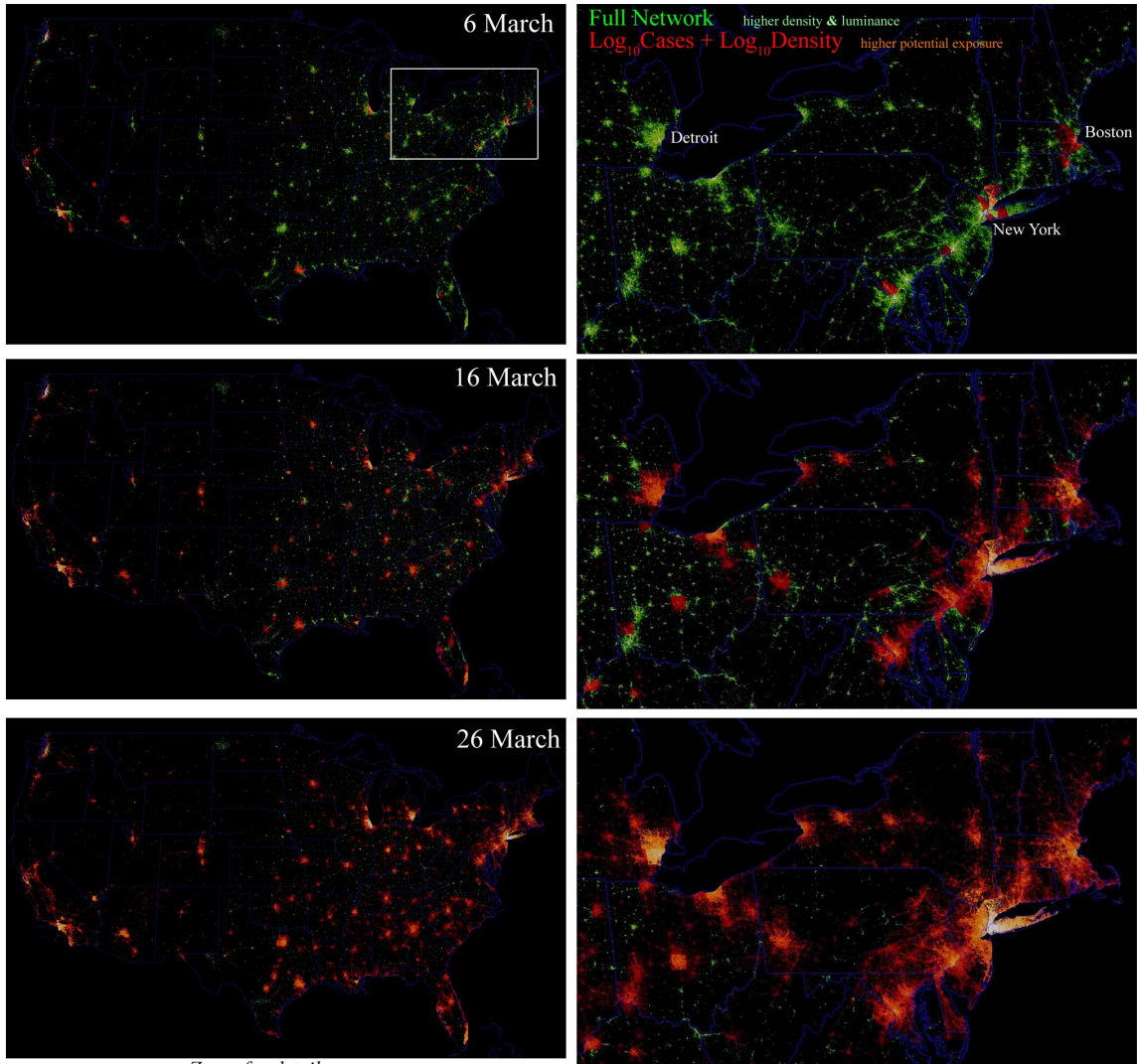

Figure 7 Propagation of SARS-CoV-2 confirmed cases through the high connectivity night light network. Despite varying county size resolution, it is clear that nearly full network penetration is achieved within the month of March. Between 22 January and 5 March, confirmed cases were limited to 6 cities in different network components and showed a very slow increase. The number of network components with confirmed cases began to increase daily after 6 March with rapid diffusion within exposed components. Daily animation at: *https://youtu.be/DA-U5XCqLO4*

Spatiotemporal transmission through the background network can be illustrated with the progression of network parameters through time. Figure 9 shows number of components, total network area and coincident population for both high and low connectivity development networks projected onto counties with confirmed cases. For each parameter, the transmission through the network was fastest between Julian days 65 and 70, although the rate of increase in confirmed cases does not begin to decrease until around Julian day 80. Air and surface mobility metrics begin to decrease only after Julian day 68, by which point transmission has encompassed almost all of the network.

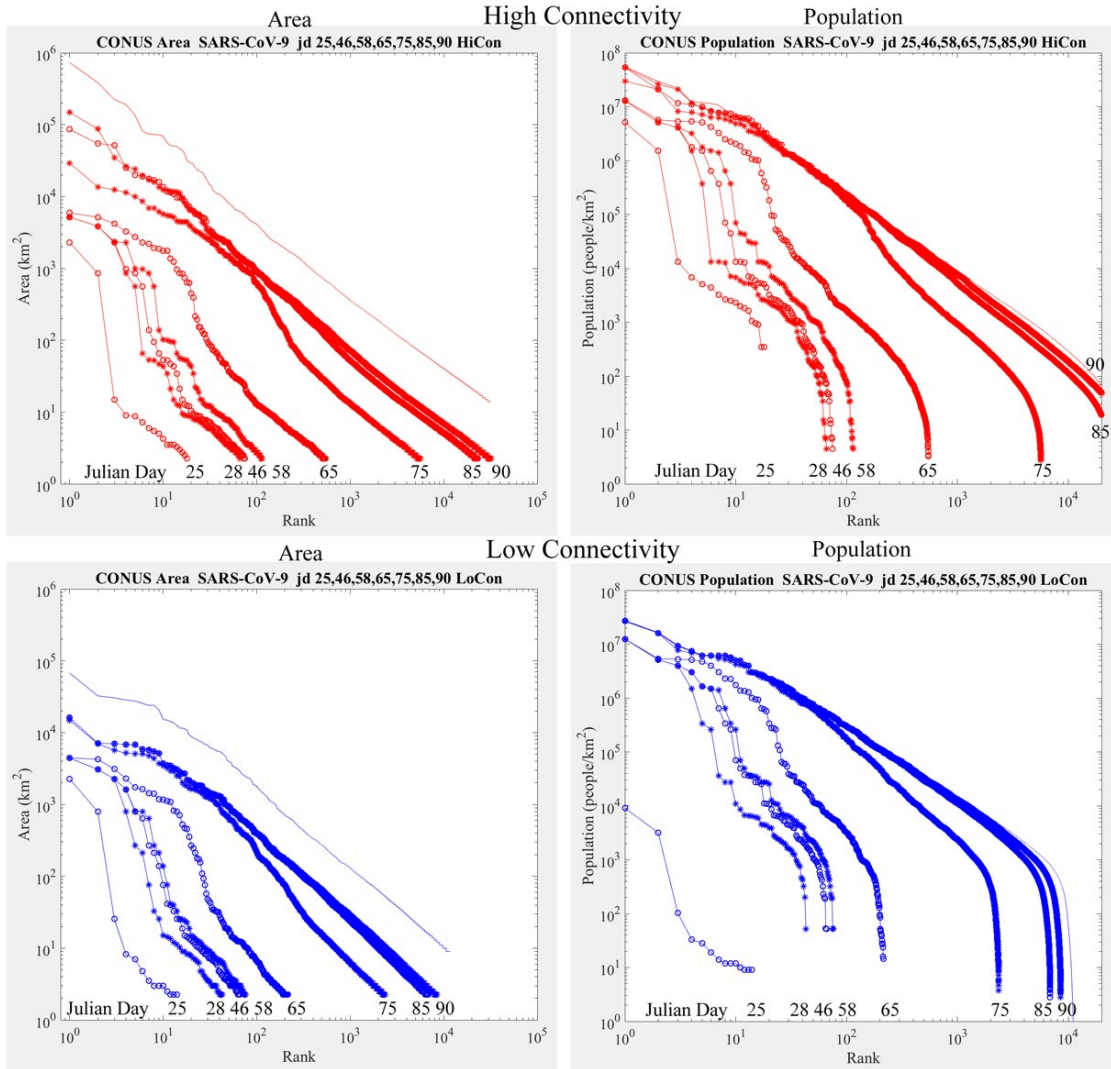

Figure 8 Propagation of SARS-CoV-2 confirmed cases through high and low connectivity night light network rank-size distributions. Prior to jd 64, only 6 network components had areas larger than $10^3$ km$^2$ with confirmed cases. After jd 64, both the number and size of the largest contiguous areas with confirmed cases grows rapidly as the wave penetrates the network, extending to larger numbers of smaller network components. By the end of March, almost all of the population is potentially exposed.

**Discussion**

As the COVID-19 pandemic progresses, epidemic peaks are occurring at variable times and with variable amplitudes in different geographic regions. The relative time for SARS-CoV-2—or any directly transmitted pathogen—to reach different population centers, as well as the intensity of epidemics within population centers, are both driven in part by connectivity of human population networks [*Balcan et al.*, 2010]. As state and local lock-down measures were adopted across much of the United States, characterizing epidemic spread using crude measures of transportation networks and vehicle traffic, or data from censuses or other administrative units, may yield inaccurate results. Using higher resolution spatial networks of population and development to define connectivity of human population networks is likely to yield more precise information for spatial epidemiological modeling of disease spread in these contexts.

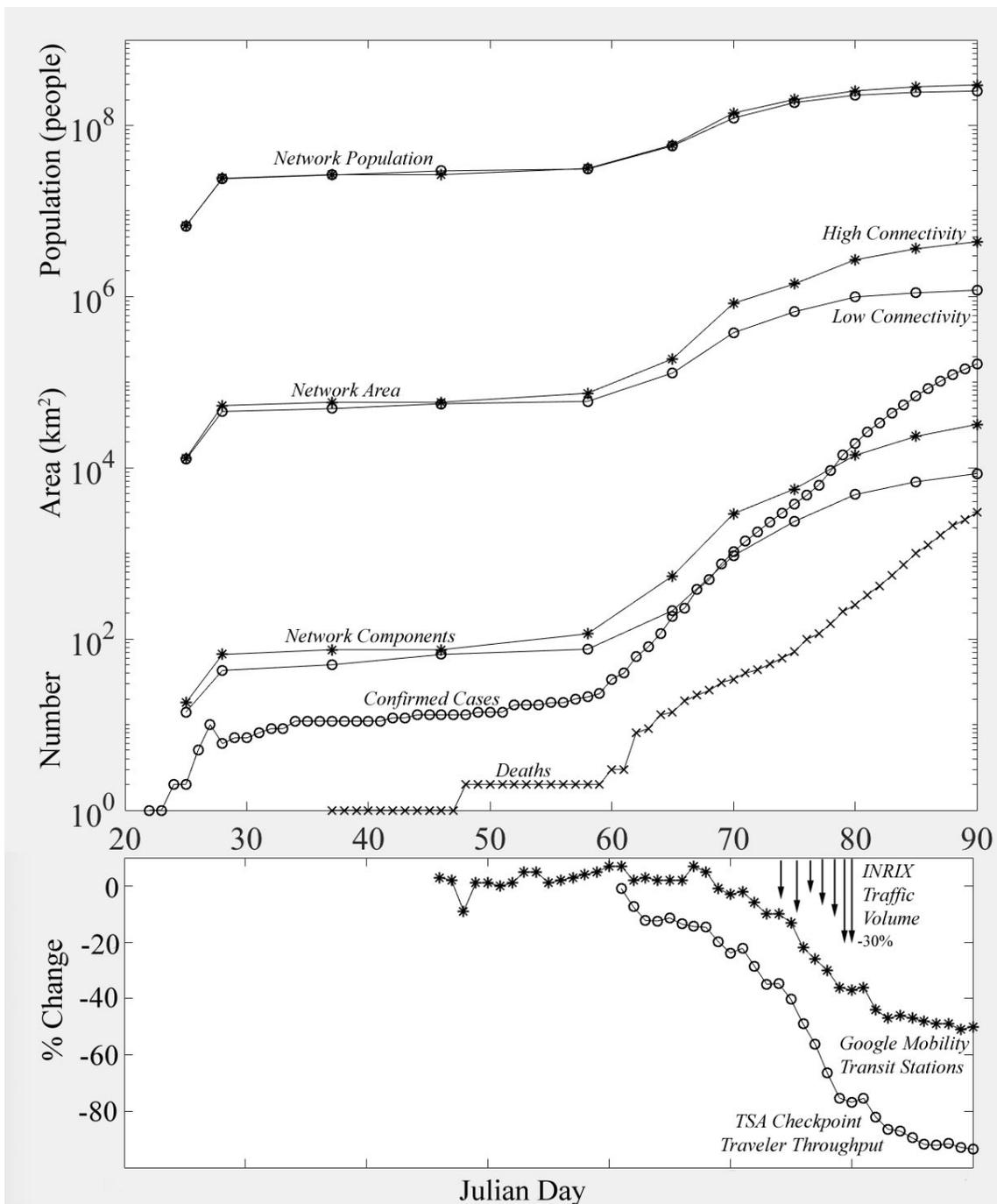

Figure 9 Network transmission timeline. Network transmission parameters with SARS-CoV-2 impacts (upper panel) and changes in mobility (lower). Comparisons of transit station visits, traffic volumes and air travelers relative to one month, three weeks and one year prior (respectively) show propagation of transmission/detection through the network had already increased rapidly by the time that mobility began to decrease around jd 69. Network population levels off before area and components because high population density components were infected earlier with propagation to lower density components occurring later. *Sources: Cases & Deaths; USAfacts.org, Traffic: Inrix.com, Transit; Google.com, Traveler; TSA.gov.*

Some projections of SARS-CoV-2 dynamics through the post-pandemic period suggest that there may be recurrent wintertime outbreaks, and that intermittent or prolonged social distancing may be necessary to avoid overwhelming healthcare systems [*Kissler et al.*, 2020]. In this context, where countries, states and municipalities are likely to relax social distancing restrictions at variable times, understanding key properties of human population networks may allow for spatially explicit prediction of the susceptibility of interconnected population centers to these 'shocks.' For example, whether and how quickly infection is likely to propagate from population centers within administrative units that relax restrictions. Such information about the connectivity of human population networks—and how schools, workplaces, hospitals or essential services are embedded within them, for example—may provide additional insight in epidemiological models employing location-specific contact matrices [*Prem et al.*, 2020] or metapopulation approaches [*Balcan et al.*, 2010], for example.

The consistency of near unity slopes of rank-size distributions spanning a wide range of network sizes has implications for community transmission within and across network components. Near unity slopes imply near uniform distributions of total component area with respect to component size interval [*Small et al.*, 2011]. If transmission begins in the largest network component(s), it may spread within infected components by person to person contact, but travel restrictions may prevent spread to uninfected network components, as illustrated by the large number of small settlements in counties without confirmed cases at the end of March 2020. However, transmission was not contained to the few components on the west coast, and eventually spread to the much larger components in the eastern and midwestern US. In animations of daily confirmed cases (*https://youtu.be/DA-U5XCqLO4*), four distinct detections occur on Julian day 63 (New York, Raleigh-Durham, Atlanta, Tampa) with a later simultaneous detections in at least 10 more components throughout the US on Julian day 68. From each nucleation point within a component, the detection of confirmed cases is appear to diffuse outward through the infected component. The variation in county size does not generally allow for inference about which specific component(s) within the county are infected and so may provide an overestimate of spatial extent. However, the progression of confirmed cases makes a clear progression from larger to smaller components for both high and low connectivity networks (Fig. 8).

We acknowledge a fundamental ambiguity in confirmed case data. It is not known to what degree these data represent disease transmission or testing frequency. However, the situation (at time of writing) with respect to confirmed case count and testing frequency (and accuracy) may be representative of a phenomenon which could occur in any scenario involving the dispersion of an entity within a spatial network. Specifically, this phenomenon is the uncertainty in attributing observations to either the *propagation of the entity itself* (in this case, the SARS-CoV-2 pathogen) or the *propagation of the detection effort* (in this case, the asynchronous, heterogeneous, and variable accuracy and availability of antigen testing throughout the US. In the general case, geospatial observations of propagation within a network can be conceptualized as a spatiotemporal convolution of two functions representing 1) propagation of the entity and 2) propagation of detection efforts. Endmember scenarios can be easily envisioned. In some cases, comprehensive detection efforts may exist prior to the beginning of transmission, and all detections could accurately represent the propagation of the entity. In the other extreme, the entity may have fully spread prior to implementation of detection efforts so observations could represent changes in detection. As the endmember scenarios are not mutually exclusive, simultaneous propagation of transmission and detection may be the most general case. In the particular case of the early propagation of SARS-CoV-2 throughout the US, we cannot speculate further until more information becomes available.

The results of the analysis reveal a number of consistencies between population and development networks.  Multiple threshold rank-size distributions of both population density and night light luminance are consistent with power laws spanning several orders of magnitude.  The spatial connectivity of the agglomerations varies with density and luminance threshold over a wide range of segment sizes – spanning 4 to 6 orders of magnitude.  These spatial networks of population and night light are considerably larger than the individual cities typically used in rank-size analyses.  The best-fit power-law domain extends through the lower tails of the distributions for night light but the lower tail roll off and the linear misfit diverges for the population distributions, as expected from underestimation in low density blocks.

While the power law provides a plausible fit to most of the distributions, the statistics say nothing about other possible distribution functions so the power law is not necessarily the only, or even the best, statistical description of the distributions.  We remain agnostic on the question of what specific distribution best describes the observations, but point out the consistency of the power law in the context of the large volume of literature assuming a power law for aggregate population distributions.  None of the conclusions we draw are conditioned on the functional form of the distributions.  We provide fits and slope estimates to quantify the linearity and implications of near-unity slope.  Area rank-size distributions are generally consistent with power laws with slopes near -1 but these distributions are a result of both the underlying process(es) and the spatial depiction by the proxy data.  Every dataset is limited by its spatial resolution as well as its accuracy.  In the case of the population data, the effective spatial resolution varies over several orders of magnitude in area of administrative unit – although the highest density areas generally have much smaller administrative units.  The light detected by VIIRS is known to extend  beyond the illuminated area because of the resolution of the sensor and atmospheric scattering of the emitted light [*Elvidge et al.*, 2007a; *Small et al.*, 2005].  This contributes somewhat to the increased connectivity at the lowest brightness levels.

**Acknowledgements**
CS gratefully acknowledges funding provided by the Fulbright Foreign Scholarship Program and the Brazilian Federal Agency for Support and Evaluation of Graduate Education. D. Sousa was funded by a postdoctoral fellowship from the La Kretz Research Center at Sedgwick Reserve.